\documentclass[rnote]{aa}

\usepackage{graphicx}
\usepackage{txfonts}
\usepackage{color}

\begin{document}

\title{Outer boundary conditions for evolving cool white
      dwarfs}
\author{Ren\'e D. Rohrmann\inst{1,2},
        Leandro G. Althaus\inst{2,3},
        Enrique Garc\'ia-Berro\inst{4,5},
        Alejandro H. C\'orsico\inst{2,3} \and
        Marcelo M. Miller Bertolami\inst{2,3}}
\institute{Instituto de Ciencias Astron\'omicas, de la Tierra y del Espacio
           (CONICET-UNSJ),
           Av. Espa\~na 1512 (sur),
           5400 San Juan, Argentina\\
           \email{rohr@icate-conicet.gob.ar}
           \and
           Member of CONICET, Argentina
           \and
           Facultad de Ciencias Astron\'omicas y Geof\'{\i}sicas,
           Universidad Nacional de La Plata,
           Paseo del Bosque s/n,
           1900 La Plata, Argentina\\
           \email{althaus@fcaglp.unlp.edu.ar}
           \and
           Departament de F\'\i sica Aplicada,
           Universitat Polit\`ecnica de Catalunya,
           c/Esteve Terrades 5,
           08860 Castelldefels, Spain\\
           \email{garcia@fa.upc.edu}
           \and
           Institute for Space Studies of Catalonia,
           c/Gran Capit\`a 2--4, Edif. Nexus 104,
           08034  Barcelona, Spain\\}
\date{\today}

\abstract{White dwarf evolution  is essentially a gravothermal cooling
           process, which, for  cool white dwarfs, sensitively depends
           on the treatment of the outer boundary conditions.}
         {We  provide detailed  outer boundary  conditions appropriate
           for computing the evolution  of cool white dwarfs employing
           detailed  non-gray  model  atmospheres  for  pure  hydrogen
           composition.  We also explore the impact on the white dwarf
           cooling times of  different assumptions for energy transfer
           in the atmosphere of cool white dwarfs.}
         {Detailed non-gray model  atmospheres are computed taken into
           account non-ideal effects in  the gas equation of state and
           chemical  equilibrium,  collision-induced  absorption  from
           molecules, and the Lyman $\alpha$ quasi-molecular opacity.}
         {We explore the impact  of outer boundary conditions provided
           by updated model atmospheres on the cooling times of $0.60$
           and $0.90\,  M_{\sun}$ white dwarf  sequences.  Our results
           show  that the  use of  detailed outer  boundary conditions
           becomes  relevant  for  effective temperatures  lower  than
           $5800$ and  $6100$~K for sequences with  $0.60$ and $0.90\,
           M_{\sun}$,   respectively.    Detailed  model   atmospheres
           predict  ages that  are  up to  $\approx  10\%$ shorter  at
           $\log(L/L_{\sun})=-4$ when  compared with the  ages derived
           using    Eddington-like   approximations    at   $\tau_{\rm
           Ross}=2/3$.   We  also   analyze  the  effects  of  various
           assumptions  and  physical processes  of  relevance in  the
           calculation of outer boundary conditions. In particular, we
           find  that  the Ly$\alpha$  red  wing  absorption does  not
           affect substantially the evolution of white dwarfs.}
         {White dwarf cooling timescales  are sensitive to the surface
           boundary      conditions      for      $T_{\rm      eff}\la
           6000$~K. Interestingly enough, non-gray effects have little
           consequences   on  these   cooling   times  at   observable
           luminosities.    In   fact,  collision-induced   absorption
           processes,  which  significantly  affect  the  spectra  and
           colors of old  white dwarfs with hydrogen-rich atmospheres,
           have not noticeable effects  in their cooling rates, except
           throughout the Rosseland mean opacity.}
\keywords{stars:  evolution  ---  stars:  interiors ---  stars:  white
          dwarfs}
\titlerunning{Boundary conditions for evolving cool white dwarfs}
\authorrunning{R. D. Rohrmann et al.}

\maketitle


\section{Introduction}
\label{intro}

An accurate assessment of the rate  at which white dwarfs cool down is
a fundamental  issue, because these  stars can be used  as independent
and accurate  age indicators.  As a  matter of fact,  white dwarfs are
the most common  end-point of stellar evolution --  see, for instance,
Althaus  et al.   (2010a)  for a  recent  review --  and  as such  are
valuable  in constraining  several  properties of  a  wide variety  of
stellar populations, including globular  and open clusters (Von Hippel
\& Gilmore  2000; Hansen et al.   2007; Winget et  al.  2009; Garc\'\i
a--Berro  et al.   2010).  Additionally,  they  can be  used to  place
constraints on elementary particles such as axions (Isern et al. 1992;
C\'orsico et al.  2001; Isern  et al.  2008), and neutrinos (Winget et
al.   2004)  or  on  alternative  theories  of  gravitation  (Garc\'\i
a--Berro et  al.  1995;  Garc\'\i a--Berro et  al.  2011).   These and
other potential applications require  a detailed and precise knowledge
of the  main physical processes  that control their evolution.   A key
ingredient   is   the  energy   transfer   in   the  atmospheric   and
sub-atmospheric  layers  which control  their  cooling (Mestel  1952).
Once  convection reaches  the outer  edge  of the  degenerate core  in
low-luminosity white  dwarfs --  the so-called convective  coupling --
the  cooling becomes  strongly  tied  to the  treatment  of the  outer
boundary  conditions  (B\"ohm  \&  Grenfell 1973).   Consequently,  an
accurate assessment  of the cooling rate at  low luminosities requires
the use  of detailed  model atmospheres (Hansen  1998; Salaris  et al.
2000; Serenelli et al. 2001).

The treatment of  the energy transfer in white  dwarf atmospheres is a
difficult  task  which  involves  the  solution of  the  equations  of
radiative transfer  coupled to convection,  in a highly  non-ideal gas
regime where  several molecular and quasi-molecular  processes have to
be considered.   This implies a  high degree of sophistication  of the
calculations, especially  at very low luminosities.  The importance of
using  detailed  boundary conditions  was  first  addressed by  Hansen
(1998, 1999).  Over the years, detailed model atmospheres that include
a  complete   treatment  of   energy  absorption  processes   such  as
collision-induced opacity (Bergeron et  al.  1991; Saumon and Jacobson
1999; Rohrmann 2001) and  the Lyman $\alpha$ wing absorption (Kowalski
\& Saumon 2006; Rohrmann et al. 2011), have been developed.

Here  we   provide  detailed   boundary  conditions  which   allow  to
consistently  compute the  evolution of  cool white  dwarfs  with pure
hydrogen atmospheres.   These boundary conditions are  provided in the
form of  tables for  a wide range  of surface gravities  and effective
temperatures.  In  the  following   sections  we  describe  the  model
atmospheres  and the evolutionary  code (\S~\ref{tools}),  and explore
the  impact  on the  cooling  times  of  different physical  processes
(\S~\ref{results}).  Conclusions are given in Sec.~\ref{conclusion}.


\section{Numerical tools}
\label{tools}

\begin{figure}[t]
\centering
\includegraphics[clip,width=220pt]{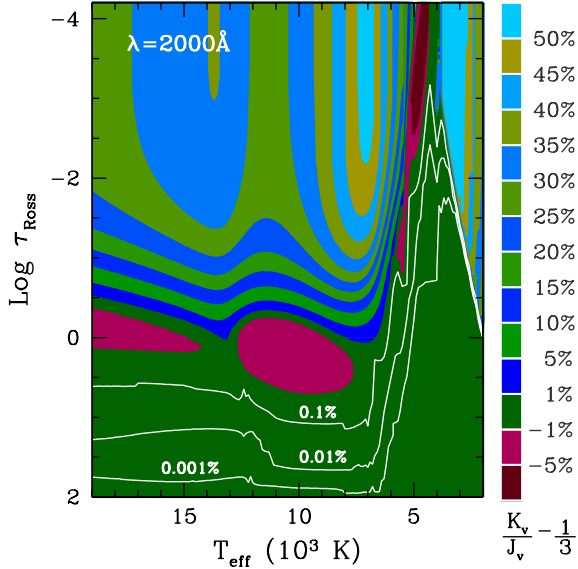}
\caption{Test  of the  diffusion approximation  in the  plane ($T_{\rm
  eff}$,  $\log \tau_{\rm  Ross}$)  based on  differences between  the
  ratio $K_\nu/J_\nu$ and its  asymptotic value $1/3$ at large optical
  depth,   for  a   wavelength   $\lambda=2000$~\AA.   These   results
  correspond to detailed non-gray models  for H atmospheres at $\log g
  =8$.}
\label{f.KJa}
\end{figure}

In our calculations, the  outer boundary conditions are obtained using
the  pure-hydrogen  LTE  model  atmospheres  described  at  length  in
Rohrmann  et   al.~(2001,  2002,  2011).    Specifically,  we  compute
pressure,  temperature, and outer  mass fraction  at an Rosseland
mean optical depth
$\tau_{\rm Ross}=25.1189$ ($\log \tau_{\rm Ross}=1.4$) for $40000~{\rm
K}\le  T_{\rm  eff}\le 2000$~K  and  $6.5\le  \log  g\le 9.5$.   Model
atmospheres were computed in the range $-6\le\log\tau_{\rm Ross}\le 2$
(in steps  of 0.1  dex) assuming hydrostatic  and radiative-convective
equilibrium.   Convective  transport   is  treated  within  the  usual
mixing-length (ML2) approximation, in which the ratio of the
mixing-length to the pressure scale height is $\alpha=1$.
The   microphysics  comprises
non-ideal  effects   in  the  gas  equation  of   state  and  chemical
equilibrium based on the  occupation probability formalism as describe
in  Rohrmann  et  al.    (2002).   The  chemical  composition  of  the
atmosphere  includes H,  H$_2$,  H$^+$, H$^-$,  H$_2^+$, H$_3^+$,  He,
He$^-$, He$^+$,  He$^{2+}$, He$_2^+$,  HeH$^+$, and e$^-$.   The level
occupation  probabilities are  self-consistently  incorporated in  the
calculation  of the  line and  continuum  opacities. Collision-induced
absorptions  due  to  H$_2$-H$_2$  (Borysow, Jorgensen  \&  Fu  2001),
H$_2$-H  (Gustafsson \&  Frommhold 2003),  H-H (Doyle  1968), H$_2$-He
(Jorgensen et al. 2000) and  H-He pairs (Gustafsson \& Frommhold 2001)
are also taken into  account. Model atmospheres explicitly include the
Lyman~$\alpha$ quasi-molecular  opacity as  a result of  eight allowed
electric dipole  transitions arising  from H-H and  H-H$_2$ collisions
(Rohrmann  et  al.~2011).   This  opacity reduces the predicted
flux at wavelength  $\lambda< 4000\AA$ for stars cooler than
$T_{\rm eff}\approx 6000$~K.

\section{Results}
\label{results}

\begin{figure}
\centering
\includegraphics[clip,width=220pt]{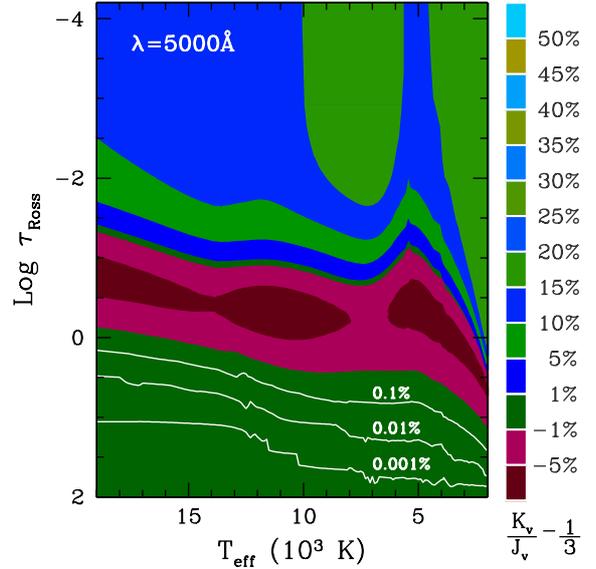}
\caption{Same as  Fig.~\ref{f.KJa}, but for $\lambda=5000\AA$.}
\label{f.KJb}
\end{figure}

The evolutionary  calculations reported here have been  done using the
{\tt LPCODE}  stellar evolutionary code  (Althaus et al.   2003, 2005,
2012).  This code has recently  been successfully used to perform very
accurate evolutionary  calculations --  see Garc\'\i a--Berro  et al.
(2010), Althaus et al.  (2010b), Renedo et al.  (2010), and references
therein.  Of relevance for this work, outer boundary condition
to the stellar structure and evolution equations are specified by
performing three envelope integrations from starting values -- as given
by the adopted  model atmosphere --  inward to a fitting outer mass
fraction, as described in Kippenhahn et al. (1967).
Energy sources resulting from crystallization -- the release
of  latent  heat and  of  energy  resulting  from carbon-oxygen  phase
separation  -- are  taken  into  account using  the  phase diagram  of
Horowitz et al.  (2010), see  Althaus et al.  (2012) for details.  The
equation  of  state  is that  of  Segretain  et  al.  (1994)  for  the
high-density  regime   --  which   accounts  for  all   the  important
contributions for  both the  liquid and solid  phases (Althaus  et al.
2007) -- complemented with an updated version of the equation of state
of Magni  \& Mazzitelli (1979) for the  low-density regime.  Radiative
opacities  are those  of  OPAL (Iglesias  \&  Rogers 1996),  including
carbon-   and   oxygen-rich   compositions,  complemented   with   the
low-temperature opacities of Ferguson et al. (2005), linearly
extrapolated to high densities when needed. Conductive
opacities are  taken from  Cassisi et al.   (2007). We also  take into
account  the  effects  of   element  diffusion  due  to  gravitational
settling,  chemical  and  thermal  diffusion of   $^1$H,  $^3$He,
  $^4$He, $^{12}$C,  $^{13}$C, $^{14}$N  and $^{16}$O, see  Althaus et
  al.  (2003)  for details.  In  particular, the metal  mass fraction
$Z$ in  the envelope of our models  is specified by scaling  it to the
local abundance  of the  CNO elements at  each layer.  To  account for
this, we  consider radiative opacities tables from  OPAL for arbitrary
metallicities.  Convection is  treated  within  the ML2  version
  ($\alpha=1$) of the mixing-length theory.

We  compute the  evolution of  sequences of  white dwarfs  of  0.6 and
$0.9\,  M_{\sun}$.   Initial  configurations  are the  result  of  the
complete   evolution  of  1.75   and  $5.0\,   M_{\sun}$  progenitors,
respectively, with  metallicity $Z=0.01$ -- see Renedo  et al.  (2010)
for details.   Progenitor stars  were evolved from  the zero  age main
sequence, through  the thermally-pulsing  and mass-loss phases  on the
asymptotic  giant branch,  to the  cooling phase.   Time-dependent
overshoot mixing  beyond the formal convective  boundary during core
hydrogen  and helium  burning stages  was taken  into  account, see
Althaus et al. (2005) for  details. Mass  loss during the RGB and AGB phases
was considered following the prescription of Schr\"oder \&  Cuntz (2005) and
Vassiliadis \& Wood (1993), respectively.  The outer chemical profiles of
our sequences are the result  of element diffusion processes that lead
to  the formation of  pure hydrogen  envelopes with  zero metallicity.
The total mass  of hydrogen left after hydrogen  burning is completely
exhausted  amounts to  $7 \times  10^{-5}$ and  $7.6  \times 10^{-6}\,
M_{\sun}$ for  the 0.6  and $0.9\, M_{\sun}$  sequences, respectively.
For each stellar mass, we computed  the cooling phase down to very low
luminosities, when most of the white dwarf has already crystallized.


Standard outer boundary conditions are usually based on the
Eddington gray approximation,  which  assumes  the  diffusion
approximation for  radiative transfer  and neglects convection  at low
optical  depths lower than  $\tau_{\rm Ross}=2/3$.
A simple test for the diffusion approximation is to compare the ratio
between the second ($K_\nu$) and zero-order ($J_\nu$) moments of the
radiation field, with its asymptotic value ($1/3$) at large depth
($\tau_{\rm  Ross}\rightarrow \infty$).
Figs.~\ref{f.KJa}  and  \ref{f.KJb} show  such  comparison at  optical
depths  $10^{-4.2}<\tau_{\rm Ross}<10^2$  for detailed,  pure hydrogen
atmospheres with  $\log g=8$ and  $2000~{\rm K}<T_{\rm eff}<19000~{\rm
K}$, and for $\lambda=2000$  and $5000\AA$, respectively.  In general,
the  difference  between $K_\nu/J_\nu$  and  $1/3$ increases  outward,
first  with negative  values and  then with  positive ones,  forming a
complex  pattern in  the plane  ($T_{\rm eff},\log  \tau_{\rm Ross}$).
Departures from  the diffusion  limit at $\tau_{\rm  Ross}=2/3$ ($\log
\tau_{\rm Ross}=-0.176$) become larger  than 1\%--5\% depending on the
wavelength and  effective temperature.  Large  deviations occur mainly
for  the visible  (Fig.~\ref{f.KJb}) and  infrared wavelengths  at low
$T_{\rm eff}$.  This depth, therefore, is not optimal to establish the
outer boundary conditions of cool white dwarfs.

\begin{figure}
\centering
\includegraphics[clip,width=220pt]{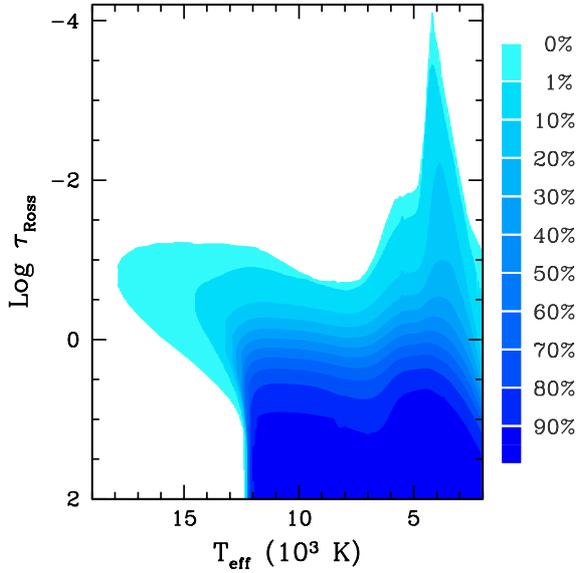}
\caption{Fraction of  convective energy in the  plane ($\log \tau_{\rm
  Ross}$, $T_{\rm eff}$) for model atmospheres with $\log g=8$.}
\label{f.conv}
\end{figure}

Convective  transport  represents another  serious  limitation of  the
standard  method to  evaluate boundary  conditions.  Fig.~\ref{f.conv}
shows the fraction  of the energy flux carried out  by convection as a
function of  $T_{\rm eff}$  in model atmospheres  with $\log  g=8$.  A
superficial  convection zone  starts at  $T_{\rm  eff}\approx 18000$~K
associated  with the  recombination of  hydrogen.  Below  $T_{\rm eff}
\approx 12200$~K,  the efficieny of convection  increases rapidly with
depth  and the  convection  zone extends  down  to the  bottom of  the
atmosphere ($\tau_{\rm Ross}=100$).  When the star cools below $T_{\rm
eff}\approx 7000$~K, the top of  the convection zone slowly extends to
very low  optical depths as  a result of H$_2$  formation.  Convection
efficiency  declines  for   models  cooler  than  $T_{\rm  eff}\approx
4000$~K.   It  is  clear  that  for $3000~{\rm  K}\la  T_{\rm  eff}\la
12500$~K, convection may carry more than 30--40\% of the total flux at
$\tau_{\rm Ross}\la 2/3$.

\begin{figure}
\includegraphics[clip,width=230pt]{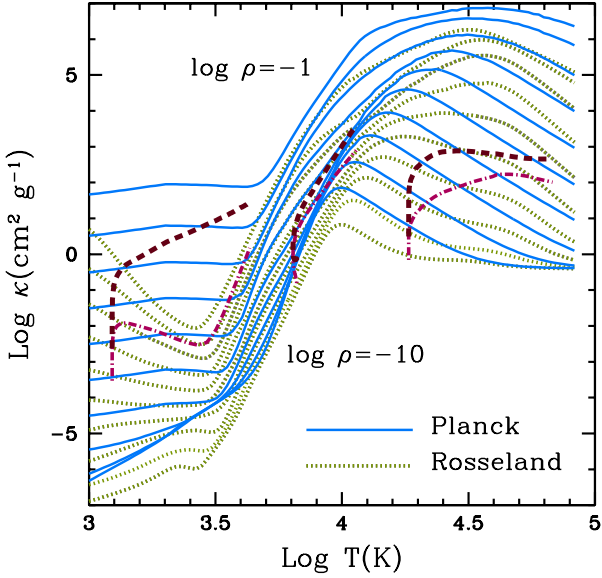}
\caption{Planck  (solid  lines)  and  Rosseland  (dotted  lines)  mean
  opacities  as  a  function  of  the temperature  at  mass  densities
  increasing from  $\log \rho=-10$ to  $\log \rho=-1$ in steps  of one
  dex (from bottom to top).  Thick dashed and dashed-dotted curves are
  the Planck and Rosseland  mean opacities for white dwarf atmospheres
  with $\log  g=8$ and $T_{\rm eff}=3000$~K  (left), $8000$~K (middle)
  and $30000$~K (right).}
\label{f.kappa2}
\end{figure}

The importance of the gray approximation throughout the atmosphere may
be  tested  comparing  two  mean  opacities  with  different  spectral
weighting,  for instance  the Planck  ($\kappa_{\rm Planck}$)  and the
Rosseland ($\kappa_{\rm  Ross}$) means  (Mihalas 1978). Note  that all
mean opacities  are equal  in a gray  atmosphere.  Fig.~\ref{f.kappa2}
displays $\kappa_{\rm Planck}$ and  $\kappa_{\rm Ross}$ for a hydrogen
gas as a function of  the temperature, for several densities.  At high
($\log T \ga 4$) and low  ($\log T \la 3.5$) temperatures, the opacity
is  mainly due to  atomic and  molecular hydrogen,  respectively, both
yielding  large   discrepancies  between  $\kappa_{\rm   Planck}$  and
$\kappa_{\rm  Ross}$, and  therefore strong  deviations from  the gray
approximation.  At intermediate temperatures ($3.5 \la \log T \la 4$),
the differences between the two mean opacities are smaller because the
H$^-$ absorption, which  is the dominant opacity source,  has a nearly
flat behavior in part  of the spectrum. Fig.~\ref{f.kappa2} also shows
the  run  of $\kappa_{\rm  Planck}$  (dashed  lines) and  $\kappa_{\rm
  Ross}$  (dash-dotted lines) for  H atmospheres  with $\log  g=8$ and
$T_{\rm eff}=3000~{\rm K}$, $8000~{\rm  K}$, and $30000~{\rm K}$ (from
left to right). As expected, the model with $T_{\rm eff}=8000$~K shows
moderate  differences  between  both  opacities ($\approx  0.5$  dex).
Discrepancies in  the hot  model ($T_{\rm eff}=30000$~K)  are somewhat
larger ($0.6-1.2$ dex), but the  largest ones ($1-3$ dex) occur in the
cooler model ($T_{\rm eff}=3000$  K), which strongly deviates from the
gray approximation.

\begin{figure}[t]
\includegraphics[clip,width=230pt]{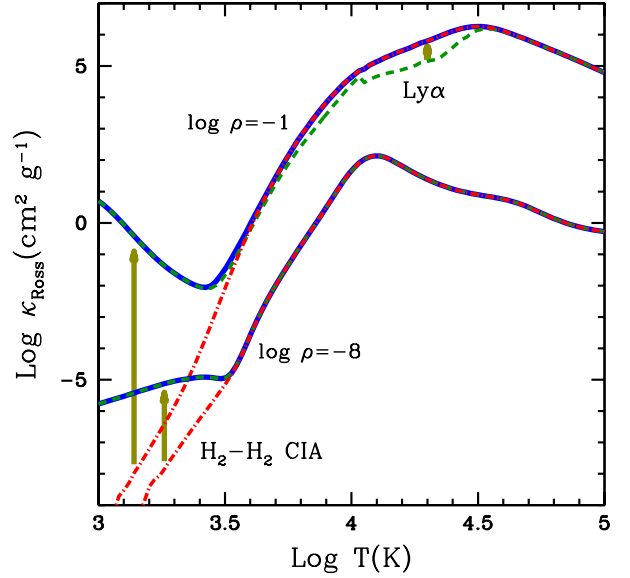}
\caption{Effects of  the collision-induced opacities  (H$_2$-H$_2$ CIA
  and  Ly$\alpha$ red  wing) on  the  Rosseland mean  opacity for  two
  densities, $\log \rho =-1$ and $\log\rho = -8$.}
\label{f.kappa1}
\end{figure}

The effects of quasi-molecular processes on the Rosseland mean opacity
are illustrated  in Fig.~\ref{f.kappa1}  for $\log \rho=-8$  and $-1$.
The  dashed   and  dashed-dotted  lines   correspond  respectively  to
calculations in which the contributions of the Ly$\alpha$ red wing and
the H$_2$-H$_2$ CIA opacities  have been removed. While the Ly$\alpha$
opacity affects the energy distribution emitted by white
dwarfs  cooler than  $T_{\rm eff}\approx  5000$~K (Kowalski  \& Saumon
2006; Rohrmann et al. 2011), it has a moderate effect on the Rosseland
opacity  at  low  densities  and  relatively  high  temperatures  (see
Fig.~\ref{f.kappa1}), when  the convective coupling  with stellar core
has  not yet occurred.   In contrast,  abrupt changes  in $\kappa_{\rm
Ross}$ occur at low temperatures due to the H$_2$-H$_2$ CIA processes,
which increase  the mean  opacity by several  orders of  magnitude and
have  important effects  on  the  cooling rates  of  old white  dwarfs
(Hansen 1999).

In  view of  the previous  remarks, it  seems important  to  treat the
boundary  conditions  as accurately  as  possible.  Detailed  non-gray
model  atmospheres  have been  computed  to  provide surface  boundary
conditions for white dwarfs with hydrogen envelopes. The values of the
pressure, temperature and outer fraction of stellar mass were obtained
at $\tau_{\rm Ross}=25.1189$ ($\log\tau_{\rm Ross}=1.4$) for effective
temperatures  ranging from  2000 to  40000~K, in  steps of  100~K, and
$\log g$ from 6.5 to 9.5, in steps of 0.1 dex.   Note  that  at
$\tau_{\rm  Ross} >  25$,  the diffusion  approximation is  guaranteed
within   0.01\%   or   better   for   most  of   the   spectrum   (see
Figs.~\ref{f.KJa} and \ref{f.KJb}).

\begin{figure}
\centering
\includegraphics[clip,width=240pt]{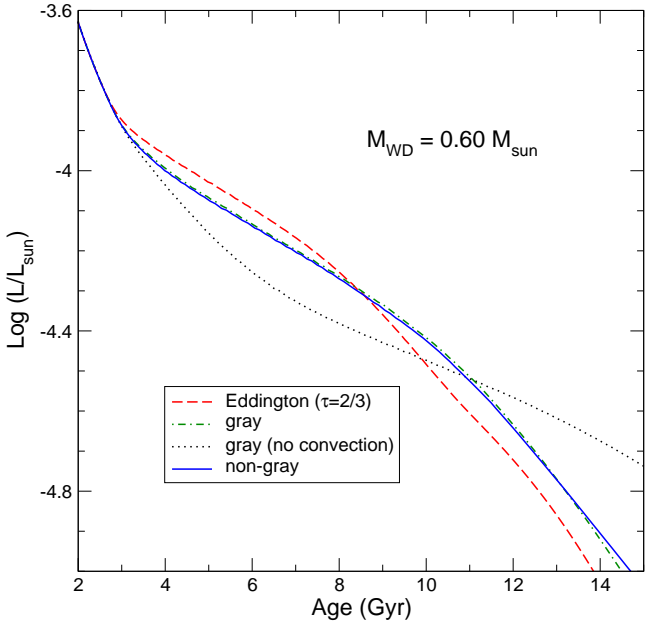}
\caption{Surface  luminosity  versus  age  for  the  $0.6\,  M_{\sun}$
  cooling  sequences   resulting  from  the  use   of  outer  boundary
  conditions as given by non-gray model atmospheres (solid blue line),
  gray model  atmospheres (dotted-dashed  green line), and  gray model
  atmospheres in which convection is neglected (dotted line).  Results
  based on  the Eddington gray  approximation are shown with  a dashed
  red line.}
\label{tcool060}
\end{figure}

\begin{figure}[t]
\centering
\includegraphics[clip,width=240pt]{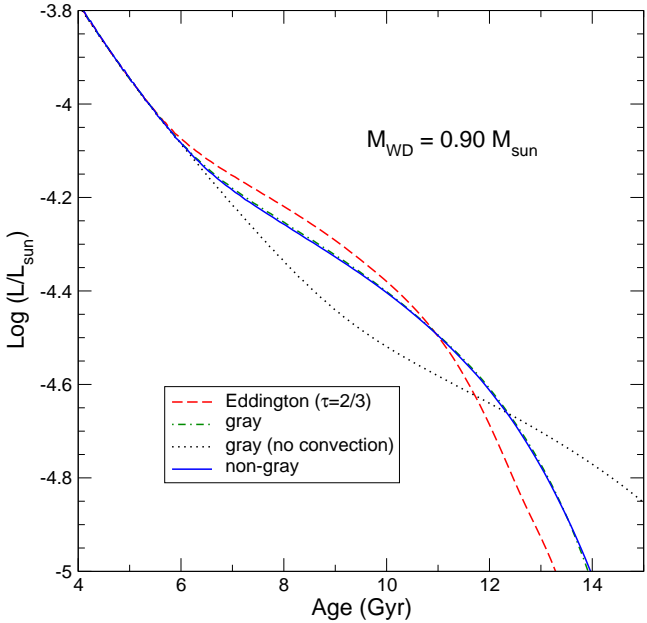}
\caption{Same  as  in  Fig.~\ref{tcool060}  for the  $0.9\,  M_{\sun}$
  sequences.}
\label{tcool090}
\end{figure}

To explore the influence of the boundary conditions on the cooling, we
display  in Figs.~\ref{tcool060}  and \ref{tcool090}  the relationship
between  the  surface  luminosity  and  age for  the  0.6  and  $0.9\,
M_{\sun}$  sequences, respectively,  that result  from  non-gray model
atmospheres (solid  line), gray model  atmosphere (dotted-dashed line)
and  gray  model atmospheres  in  which  the  convection is  neglected
(dotted  line).  These figures  also show  results obtained with
boundary conditions at $\tau_{\rm Ross}=2/3$ based on the Eddington
gray approximation (dashed
line). Clearly, the use  of detailed outer boundary conditions becomes
relevant for cooling once evolution has proceeded to luminosities
lower  than  $\log(L/L_{\sun})\sim-3.8$  ($-4.0$)  in  the  case  of
$0.60\,  M_{\sun}$  ($0.90\,  M_{\sun}$) models.   These  luminosities
correspond to effective temperatures  lower than 5800~K (6100~K). Such
values indicate the onset of the convective coupling between the outer
envelope  and the  isothermal  degenerate core  (Tassoul et al. 1990;
D'Antona \& Mazzitelli  1990; Prada Moroni  \& Straniero
2007).   For larger  effective temperatures, evolution  is almost
insensitive to a detailed treatment of the outer boundary conditions.

Figs.~\ref{tcool060} and \ref{tcool090} show  that the use of detailed
model atmospheres directly translates  into cooling times different from
those predicted by the standard Eddington approximation.
Note that, for the $0.6\, M_{\sun}$ sequence, in
the range $-3.8\ga \log(L/L_{\sun})\ga  -4.3$, the inclusion of proper
outer boundary conditions decreases the cooling ages by up to 0.7 Gyr,
while  this trend  in the  cooling times  is reversed  at luminosities
below  $\log(L/L_{\sun})\sim-4.3$, where the  use of  detailed model
atmospheres  results in  longer cooling  times.  The  behavior  of the
cooling  times   is  qualitatively   similar  for  the   more  massive
sequence.  Again, the  use  of any  Eddington-like approximation  that
involves the diffusion assumption  for radiative transfer and neglects
convection at  low optical depths, incorrectly  predicts the evolution
of cool white dwarfs.

\begin{figure}[t]
\centering
\includegraphics[clip,width=240pt]{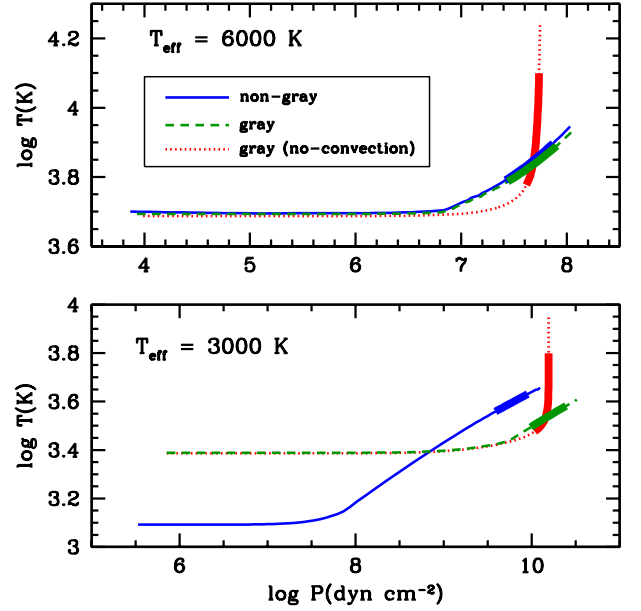}
\caption{Temperature-pressure   stratifications   of   pure   hydrogen
  atmospheres with  $T_{\rm eff}=3000$~K and 6000~K  ($\log g=8$), for
  different assumptions as  indicated on the plot. The  top and bottom
  layers of  each model are  located at $\tau_{\rm  Ross}=10^{-6}$ and
  $\tau_{\rm Ross}=100$, respectively. The thick lines show the layers
  located   between   $\tau_{\rm   Ross}\approx2/3$   and   $\tau_{\rm
  Ross}\approx25$.}
\label{tp}
\end{figure}

Surprisingly,    and   contrary    to   what    was    expected   from
Fig.~\ref{f.kappa2},  Figs.~\ref{tcool060} and \ref{tcool090}  show no
appreciable changes  in the cooling times when  the gray approximation
is  assumed  in  the  model  atmospheres. In  these  calculations  the
monochromatic opacity coefficient was forced  to take the value of the
Rosseland mean opacity. In fact, Fig.~\ref{tp} shows that at the onset
of   convective    coupling   ($T_{\rm   eff}\approx    6000$~K)   the
temperature-pressure stratifications  of gray and  non-gray models are
practically identical.  Below $T_{\rm eff}\approx 5000$~K, H$_2$-H$_2$
collision-induced   opacity  reduces   the  surface   temperature  and
increases the  temperature in deep  atmospheric layers respect  to the
gray   model.   These  non-gray   effects  increase   towards  $T_{\rm
eff}\approx  3000$~K  (Fig.~\ref{tp}),   which  corresponds  to  about
$\log(L/L_{\sun})\sim-4.95$  for the  $0.6\,M_{\sun}$  sequence (see
Fig.~\ref{tcool060}), but nevertheless have little consequences on the
cooling times at observable  luminosities.  Indeed, they are prominent
at very low  $T_{\rm eff}$, and are expected  to influence the cooling
times  once  evolution  has   proceeded  to  luminosities  lower  than
$\log(L/L_{\sun})\sim-4.95$.    We   have   also  computed   cooling
sequences  where the  collision-induced broadening  of  Ly$\alpha$ was
omitted in  model atmospheres, and  we found that this  opacity source
does not affect the  evolution substantially.  These results show that
processes  which markedly  alter the  distribution of  spectral energy
radiated by the star may have no effect on its cooling time.

On  the  contrary,  neglecting  convective energy  transfer  in  model
atmospheres  strongly alters  the  cooling times.   As illustrated  in
Fig.~\ref{tp},   neglecting   convection   results  in   much   larger
temperatures  at the  base  of  the atmosphere  (dotted  lines in  the
figure), thus  producing a  markedly shallower outer  convection zone,
and  eventually resulting  in age  differences of  up to  2  Gyr.   In
particular, for a $0.6\, M_{\sun}$ model at $\log(L/L_{\sun})=-4$, the
use of  outer boundary conditions derived from  model atmospheres that
neglect convection down to an optical depth of $\tau_{\rm Ross}\approx
25$  leads  to  an   outer  convective  zone  with  mass  $\log(M_{\rm
conv}/M_{\rm WD})\sim-18$, whereas the value resulting when convection
is considered is $-13.5$.

\begin{figure}[t]
\centering
\includegraphics[clip,width=240pt]{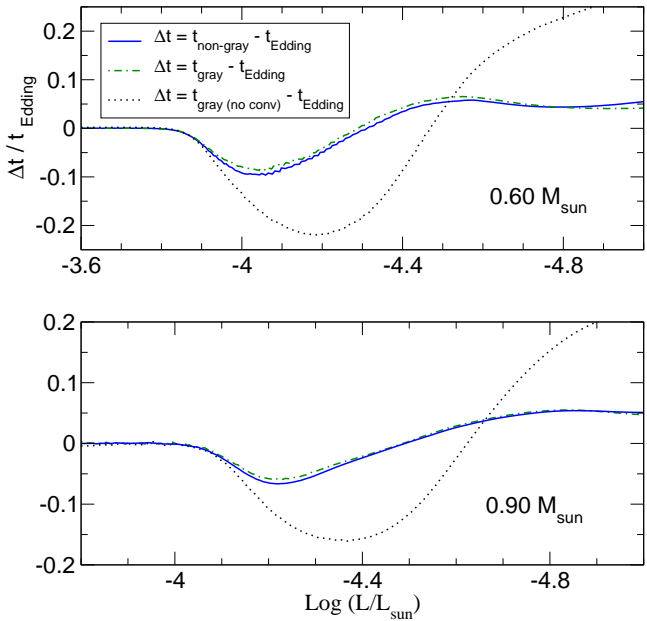}
\caption{Age  differences  between  sequences that  consider  non-gray
  model  atmospheres   (solid  blue  line),   gray  model  atmospheres
  (dotted-dashed  green line),  and  gray model  atmospheres in  which
  convection is  neglected (dotted line)  with respect to the  case in
  which the Eddington gray  approximation is considered. The upper and
  bottom  panel  correspond, respectively,  to  the  $0.6$ and  $0.9\,
  M_{\sun}$ white dwarf sequences.}
\label{dif}
\end{figure}

The impact on  the cooling times of the  different boundary conditions
can  be   better  appreciated  by   inspecting  Fig.~\ref{dif},  which
illustrates the age differences with  respect to the case in which the
Eddington  gray  approximation is  used  for  sequences that  consider
non-gray model  atmospheres (solid blue line),  gray model atmospheres
(dotted-dashed  green  line),  and  gray model  atmospheres  in  which
convection  is neglected (dotted  line).  The  upper and  bottom panel
correspond   to   the    $0.6$   and   $0.9\,   M_{\sun}$   sequences,
respectively.  Note  that  for   the  $0.6\,  M_{\sun}$  sequence  age
differences  are negative  in the  range  $-3.8\ga \log(L/L_{\sun})\ga
-4.3$, and the  use  of  detailed model  atmospheres
predicts   ages  that   are  up   to  $\approx   10  \%$   shorter  at
$\log(L/L_{\sun})=-4.05$ when compared with the ages derived using the
Eddington approximation.  The differences are somewhat smaller for the
$0.9\, M_{\sun}$ sequence, reaching up to $\approx 7\%$ at
$\log(L/L_{\sun})= -4.2$. Besides, the gray assumption translates into
age differences less than $\sim 1\%$ in both sequences.

Since the input physics  adopted in the  codes used to compute the
stellar interior and the  atmosphere is not exactly the same, we have
examined the impact of matching atmosphere and interior models at different
optical depths.
We found that the cooling times differ at most $0.7$\% ($0.9$\%) for
boundary conditions at $\tau_{\rm Ross}\approx 50$ ($100$) respect
to those obtained at $\tau_{\rm Ross}\approx 25$. Thus, differences in the
constitutive physics in the  codes appear to have small consequences
in the derivation of the boundary conditions.
We have also tested the effects of changing the value of the mixing length
parameter ($\alpha$) in the convection theory. In particular,
if the efficiency of convection is increased to $\alpha=2$ (ML3 version of
the mixing-lenght theory), the relative age differences respect to the use
of $\alpha=1$ (ML2 version) becomes smaller than $0.8$\% ($0.4$\%) for the
$0.6\,M_{\sun}$ ($0.9\,M_{\sun}$) model.


\section{Conclusions}
\label{conclusion}

The purpose of  this work has been to  provide detailed outer boundary
conditions  which  allow  to   compute  white  dwarf  evolution  in  a
consistent  way with  the predictions  of detailed  model atmospheres.
Data are  provided in the  form of tables  for a
wide range of surface  gravities and effective temperatures, which are
appropriate for computing the evolution of cool white dwarfs with pure
hydrogen  atmospheres.   The  full  set   of  data  is   available  at
http://www.icate-conicet.gob.ar/rohrmann/tables.html  or  upon request
to the authors at their e-mail addresses.

White dwarf  cooling timescales are sensitive to  the surface boundary
conditions for  effective temperatures lower  than $T_{\rm eff}\approx
6000$~K.    Different  outer   boundary  conditions   may   result  in
substantial differences in the  cooling times for cooler white dwarfs.
However, non-gray effects do not become important in the cooling rates.
On the other hand, depending on the stellar luminosity, the use of
detailed model atmospheres like the ones presented here results in age
differences of  about $  10 \%$ when  compared with the  ages computed
using  the  Eddington   approximation,  which  assumes  the  diffusion
approximation for  radiative transfer  and neglects convection  at low
optical  depths lower than  $\tau_{\rm Ross}=2/3$.   These differences
are  on the  order of  the current  uncertainties in  the  white dwarf
cooling times  at low luminosities  that result from  uncertainties in
the treatment  of progenitor  evolution, particularly during  the core
helium  burning phase  (Prada  Moroni \&  Straniero  2002; Salaris  et
al. 2010).  Consequently,  accurate outer boundary conditions provided
by detailed  model atmospheres have  to be considered  in evolutionary
studies aimed at using these stars as accurate cosmic clocks.


\begin{acknowledgements}
This  research   was  supported   by  PIP  112-200801-01474   and  PIP
112-200801-00940 from  CONICET, by MCINN grant  AYA2011--23102, by the
European Union FEDER funds, and  by the ESF EUROGENESIS project (grant
EUI2009-04167)
\end{acknowledgements}


\end{document}